\documentclass{ws-mpla}
\usepackage{hyperref}

\begin{document}

\markboth{Lixin Xu and Hongya Liu, Baorong Chang}{Correspondence
Between $5D$ Ricci-Flat Cosmological Models and Quintessence Dark
Energy Models}

\title{Correspondence Between $5D$ Ricci-Flat Cosmological Models and Quintessence Dark Energy Models}
\author{Lixin Xu\footnote{lxxu@dl.cn}, Hongya Liu\footnote{
Corresponding author: hyliu@dlut.edu.cn}, Baorong Chang}
\address{Department of Physics, Dalian University of
Technology, Dalian, 116024, P. R. China}

\maketitle

\pub{Received 19 May 2005}{Revised (Day Month Year)}

\begin{abstract}
We study the accelerating expansion and the induced dark energy of
the $5D$ Ricci-flat universe which is characterized by having a
big bounce as opposed to a big bang. We show that the arbitrary
function $\mu(t)$ contained in the $5D$ solutions can be rewritten
in terms of the redshift $z$ as a new arbitrary function $f(z)$,
and we find that there is a correspondence between this $f(z)$ and
the potential $V(\phi)$ of the $4D$ quintessence models. Using
this correspondence, the arbitrary function $f(z)$ and the $5D$
solution could be specified for a given form of the potential
$V(\phi)$.

\end{abstract}

%
\catchline{}{}{}{}{} %

\ccode{PACS Nos.: 04.50.+h, 98.80.-k.}

\section{Introduction}\label{introduction}

Recent observations of high redshift Type Ia supernovae reveal
that our universe is undergoing an accelerated expansion rather
than decelerated expansion \cite{RS,TKB,Riess}. In addition, the
discovery of Cosmic Microwave Background (CMB) anisotropy on
degree scales together with the galaxy redshift surveys indicate
$\Omega _{total}\simeq 1$ \cite{BHS} and $\Omega _{m}\simeq \left.
1\right/ 3$. All these results strongly suggest that the universe
is permeated smoothly by 'dark energy' which has a negative
pressure and violates the strong energy condition. The dark energy
and accelerating universe has been discussed extensively from
different points of view \cite{Quintessence,Phantom,K-essence}. In
principle, a natural candidate to dark energy could be a small
cosmological constant. However, there exist serious theoretical
problems: fine tuning problem and coincidence problem. To overcome
the coincidence problem, some self-interacting scaler fields
$\phi$ with an equation of state (EOS) $w_{\phi}=p_{\phi}\left/
\rho_{\phi}\right.$ were introduced dubbed quintessence, where
$w_{\phi}$ is time varying and negative. In principle, the
potentials of the scalar field would be determined from the
underlying physical theories, such as Supergravity,
Superstring/M-theory etc.. However, disregarding these underlying
physical theories and just starting from the phenomenal level, one
can design many kinds of potentials to solve the concrete problems
\cite{Quintessence,sahni}. Once the potentials are given, EOS
$w_{\phi}$ of dark energy can be found. On the contrary, the
potential can be reconstructed by a given EOS $w_{\phi}$
\cite{GOZ}. Then, the forms of the scalar potential can be
obtained by the observations data with a given EOS $w_{\phi}$.

The idea that our world may have more than four dimensions is due
to Kaluza \cite{Kaluza}, who unified Einstein's theory of General
Relativity with Maxwell's theory of Electromagnetism in a $5D$
manifold. In 1926, Klein reconsidered Kaluza's idea and treated
the extra dimension as a compacted small circle topologically
\cite{Klein}. Afterwards, the Kaluza-Klein idea has been studied
extensively from different points of view. Among them, a kind of
theory named Space-Time-Matter theory is designed to incorporate
the geometry and matter by Wesson and his collaborators, for
reviews please see \cite{Wesson} and references therein. In STM
theory, our world is a hypersurface embedded in a five-dimensional
Ricci flat ($R_{AB}=0$) manifold and all the matter in our world
are induced from the higher dimension, which is supported by
Campbell's theorem \cite{Compbell} which says that any analytical
solution of Einstein field equation of $N$ dimensions can be
locally embedded in a Ricci-flat manifold of $\left(N+1 \right)$
dimensions. Since the matter are induced from the extra dimension,
this theory is also called induced matter theory. The application
of the idea about induced matter or induced geometry can also be
found in other situations \cite{FSS}. The STM theory allows the
metric components to be dependent on the extra dimension and does
not require the extra dimension to be compact or not. The
consequent cosmology in STM theory is studied in \cite{LiuW},
\cite{STM-cosmology}, \cite{WLX}.

\section{$4D$ dark energy models with quintessence}

In a $4D$ spatially flat FLRW universe, the Friedmann equation can
be written as
\begin{equation}
H^2=\frac{8\pi
G}{3}\left(\Sigma_{i}\rho_{\gamma_{i}}+\rho_{\phi}\right),\label{4Fridemann}
\end{equation}
where, the universe is dominated by barotropic perfect fluids with
equation of state (EOS) $\rho_{\gamma_{i}}=\gamma_{i}
p_{\gamma_{i}}$ ($0\le \gamma_{i} \le 1$) (where $\gamma=0$ for
pressureless cold dark matter and $\gamma=1/3$ for radiation) and
spatially homogenous scalar filed $\phi$, dubbed quintessence.

The energy density and pressure of the scalar field $\phi$ are
\begin{eqnarray}
\rho_{\phi}&=&\frac{1}{2}\dot{\phi}^2+V(\phi),\nonumber \\
p_{\phi}&=&\frac{1}{2}\dot{\phi}^2-V(\phi),\label{scalarfield}
\end{eqnarray}
respectively, where $V(\phi)$ is the potential of the scalar
field. The potential of the scalar fields is designed in different
forms to obtain desired properties as has been interpreted in the
introduction \ref{introduction}. The equation of state of the
scalar field is
\begin{equation}
w_{\phi}=\frac{p_{\phi}}{\rho_{\phi}}.\label{eosscalar}
\end{equation}
Using Eqs. (\ref{scalarfield})-(\ref{eosscalar}), one has
\begin{eqnarray}
\frac{1}{2}\dot{\phi}^2&=&\frac{1}{2}\left(1+w_{\phi}\right)\rho_{\phi},\nonumber\\
V(\phi)&=&\frac{1}{2}\left(1-w_{\phi}\right)\rho_{\phi}.
\end{eqnarray}
The evolution equation of the scalar field $\phi$ is
\begin{equation}
\dot{\rho_{\phi}}+3H\left(\rho_{\phi}+p_{\phi}\right)=0,
\end{equation}
which yields \cite{GOZ}
\begin{eqnarray}
\rho_{\phi}(z)&=&\rho_{\phi
0}\exp\left[3\int_{0}^{z}(1+w_{\phi})d\ln(1+z)\right]\nonumber\\
&\equiv&\rho_{\phi 0}E(z),
\end{eqnarray}
where, $1+z=\frac{a_0}{a}$ is the redshift and the subscript $0$
denotes the current value. Then one obtains the potential in term
of $w_{\phi}$
\begin{equation}
V[(\phi)]=\frac{1}{2}(1-w_{\phi})\rho_{\phi 0}E(z).\label{Vz}
\end{equation}
In terms of redshift, the Friedmann equation can be written as
\begin{equation}
H(z)^2=H_0^2\left[\Sigma_{i}\Omega_{\gamma_{i}
0}(1+z)^{3(\gamma_{i}+1)}+\Omega_{\phi 0}E(z)\right],\label{Hz}
\end{equation}
where, $\Omega_{\gamma_{i} 0}$s and $\Omega_{\phi 0}$ are the
current values of dimensionless density parameters determined by
observations.

\section{Dark energy in 5D models}\label{DE5}

Within the framework of STM theory, a class of exact $5D$
cosmological solution was given by Liu and Mashhoon in 1995
\cite{Liu}. Then, in 2001, Liu and Wesson \cite{LiuW} restudied
the solution and showed that it describes a cosmological model
with a big bounce as opposed to a big bang. The $5D$ metric of
this solution reads
\begin{equation}
dS^{2}=B^{2}dt^{2}-A^{2}\left(
\frac{dr^{2}}{1-kr^{2}}+r^{2}d\Omega ^{2}\right) -dy^{2}
\label{5-metric}
\end{equation}
where $d\Omega ^{2}\equiv \left( d\theta ^{2}+\sin ^{2}\theta
d\phi ^{2}\right) $ and
\begin{eqnarray}
A^{2} &=&\left( \mu ^{2}+k\right) y^{2}+2\nu y+\frac{\nu
^{2}+K}{\mu ^{2}+k},
\nonumber \\
B &=&\frac{1}{\mu }\frac{\partial A}{\partial t}\equiv
\frac{\dot{A}}{\mu }. \label{A-B}
\end{eqnarray}
Here $\mu =\mu (t)$ and $\nu =\nu (t)$ are two arbitrary functions
of $t$, $k$ is the $3D$ curvature index $\left(k=\pm 1,0\right)$,
and $K$ is a constant. This solution satisfies the 5D vacuum
equation $R_{AB}=0$. So, the three invariants are
\begin{eqnarray}
I_{1} &\equiv &R=0, I_{2}\equiv R^{AB}R_{AB}=0,
\nonumber  \\
I_{3} &=&R_{ABCD}R^{ABCD}=\frac{72K^{2}}{A^{8}}. \label{3-invar}
\end{eqnarray}
The invariant $I_{3}$ in Eq. (\ref{3-invar}) shows that $K$
determines the curvature of the 5D manifold. It would be pointed
out that the $5D$ and $4D$ Planck mass are not related directly,
because in the STM theory one always has $^{5}G_{AB}=0$. So, in
$4D$ we can take $\kappa_4^2=8\pi G_4$, where $G_4$ is $4D$
Newtonian gravitation constant.

Using the $4D$ part of the $5D$ metric (\ref{5-metric}) to
calculate the $4D$ Einstein tensor, one obtains
\begin{eqnarray}
^{(4)}G_{0}^{0} &=&\frac{3\left( \mu ^{2}+k\right) }{A^{2}},
\nonumber \\
^{(4)}G_{1}^{1} &=&^{(4)}G_{2}^{2}=^{(4)}G_{3}^{3}=\frac{2\mu \dot{\mu}}{A%
\dot{A}}+\frac{\mu ^{2}+k}{A^{2}}.  \label{einstein}
\end{eqnarray}
In our previous work \cite{WLX}, the induced matter was set to be
a conventional matter plus a time variable cosmological `constant'
or three components: dark matter radiation and $x$-matter. In this
paper, we assume that the induced matter contains four parts: CDM
$\rho_{cd}$, baryons $\rho_b$, radiation $\rho_r$ and dark energy
$\rho_{de}$. So, we have
\begin{eqnarray}
\frac{3\left( \mu ^{2}+k\right) }{A^{2}} &=&\rho_{cd}+\rho_b+\rho_r+\rho_{de},  \nonumber \\
\frac{2\mu \dot{\mu}}{A\dot{A}}+\frac{\mu ^{2}+k}{A^{2}}
&=&-\left(p_{cd}+p_b+p_r+p_{de}\right), \label{FRW-Eq}
\end{eqnarray}
where
\begin{equation}
p_{cd}=0,\quad p_b=0, \quad p_r=\rho_r/3, \quad
p_{de}=w_{de}\rho_{de}. \label{EOS-X}
\end{equation}
From Eqs.(\ref{FRW-Eq}) and (\ref{EOS-X}), one obtains the EOS of
the dark energy
\begin{equation}
w_{de}=\frac{p_{de}}{\rho_{de}}=-\frac{2\left. \mu
\dot{\mu}\right/ A \dot{A}+\left. \left( \mu ^{2}+k\right) \right/
A^{2}+\rho_{r0}A^{-4}/3}{3\left. \left( \mu ^{2}+k\right) \right/
A^{2}-\rho_{cd0}A^{-3}-\rho_{b0}A^{-3}-\rho_{r0}A^{-4}},\label{wx}
\end{equation}
and the dimensionless density parameters
\begin{eqnarray}
\Omega _{cd} &=&\frac{\rho
_{cd}}{\rho_{cd}+\rho_{b}+\rho_{r}+\rho_{de}}=\frac{\rho
_{cd0}}{3\left( \mu ^{2}+k\right) A},  \label{omega-cd} \\
\Omega _{b} &=&\frac{\rho
_{b}}{\rho_{cd}+\rho_{b}+\rho_{r}+\rho_{de}}=\frac{\rho
_{b0}}{3\left( \mu ^{2}+k\right) A},  \label{omega-b} \\
\Omega _{r} &=&\frac{\rho
_{r}}{\rho_{cd}+\rho_{b}+\rho_{r}+\rho_{de}}=\frac{\rho
_{r0}}{3\left( \mu ^{2}+k\right) A^2},  \label{omega-r} \\
\Omega_{de} &=&1-\Omega_{cd}-\Omega_{b}-\Omega_{r}.
\label{omega-de}
\end{eqnarray}
where $\rho _{cd0}=\bar{\rho}_{cd0}A_{0}^{3}$, $\rho
_{b0}=\bar{\rho}_{b0}A_{0}^{3}$ and $\rho
_{r0}=\bar{\rho}_{r0}A_{0}^{4}$. The Hubble and deceleration
parameters were given in \cite{LiuW}, \cite{WLX},
\begin{eqnarray}
H&\equiv&\frac{\dot{A}}{A B}=\frac{\mu}{A} \\
q \left(t, y\right)&\equiv&\left.
-A\frac{d^{2}A}{d\tau^{2}}\right/\left(\frac{dA}{d\tau}\right)^{2}
=-\frac{A \dot{\mu}}{\mu \dot{A}}, \label{df}
\end{eqnarray}
from which we see that $\dot{\mu}\left/\mu\right.>0$ represents an
accelerating universe, $\dot{\mu}\left/\mu\right.<0$ represents a
decelerating universe. So the function $\mu(t)$ plays a crucial
role in defining the properties of the universe at late time.

\section{Late time evolution of the cosmological parameters versus redshift}\label{P}

We only consider the spatial flat case $k=0$. In Eqs.
(\ref{FRW-Eq})-(\ref{df}), $\nu(t)$ does not explicitly appear in
the equations. So, to avoid boring to choose the concrete forms
$\nu(t)$, we use $A_{0}\left/A \right.=1+z$ and define
$\mu_{0}^{2}\left/ \mu_{z}^{2}\right.=f\left(z\right)$ (noting
that $f(0)\equiv 1$), and then we find that the Eqs.
(\ref{wx})-(\ref{df}) reduce to
\begin{eqnarray}
w_{x} &=&-\frac{1+\Omega_{r}+\left(1+z\right)d\ln
f\left(z\right)\left/dz\right.}{3-3\Omega_{cd}
-3\Omega_{b}-3\Omega_{r}}, \label{wx-2} \\
\Omega_{cd}&=&\Omega_{cd0}\left(1+z\right)f\left(z\right),\label{omega-cd-2} \\
\Omega_{b} &=&\Omega_{b0}\left(1+z\right)f\left(z\right),\label{omega-b-2}\\
\Omega_{r} &=&\Omega_{r0}\left(1+z\right)^2 f\left(z\right),\label{omega-r-2}\\
\Omega_{de}&=&1-\Omega_{cd}-\Omega_{b}-\Omega_{r},\label{omega-de-2}\\
q&=&\frac{1+3\Omega_{x}w_{x}+\Omega_r}{2}=-\frac{\left(1+z\right)}{2}\frac{d\ln
f\left(z\right)}{dz}. \label{q}
\end{eqnarray}
As is known from the quintessence and phantom dynamical dark
energy models, there exist undefined potentials $V(\phi)$. One can
choose different forms of the potential $V(\phi)$ to describe the
desired properties of dark energy \cite{sahni}, in which many
forms of the potential are given. Now, there is an arbitrary
function $\mu(t)$ in the present $5D$ model. Different choice of
$\mu(t)$ corresponds to different choice of the potential
$V(\phi)$ in quintessence or phantom models. By transforming $t$
to redshift $z$, the choices of $\mu(t)$ corresponds to the choice
of $f(z)$. This enables us to look for desired properties of the
universe via Eqs. (\ref{wx-2})-(\ref{q}). In these definition, the
Friedmann equation becomes
\begin{equation}
H^2=H_0^2(1+z)^2f(z)^{-1}, \label{friedmann}
\end{equation}
where, $H_0$ is the current Hubble parameter. These enable us to
use the supernova observations data to constrain the parameters
contained in the model or function $f(z)$. And then we can
postulate it in the early evolution epoch of the universe and
discuss the possible phenomena. By comparing Eq. (\ref{friedmann})
with Eq. (\ref{Hz}), we find that there exists some correspondence
between the undefined potential $V(\phi)$ and the function $f(z)$.
In naive, we can take $f(z)$ as follows
\begin{equation}
f(z)=(1+z)^2\left[\Sigma_{i}\Omega_{\gamma_{i}
0}(1+z)^{3(\gamma_{i}+1)}+\Omega_{\phi
0}E(z)\right]^{-1},\label{fz}
\end{equation}
From Eq. (\ref{fz}), it is easy to see that the function $E(z)$ is
determined by the particular potential $V(\phi)$ and by this kind
of correspondence the function $f(z)$ is defined. Then, the
evolutions of the density components and the EOS of dark energy
can be derived in this way. As did in Ref. \cite{GOZ}, we consider
the following cases as examples

\begin{figure}
\centering
\includegraphics[]{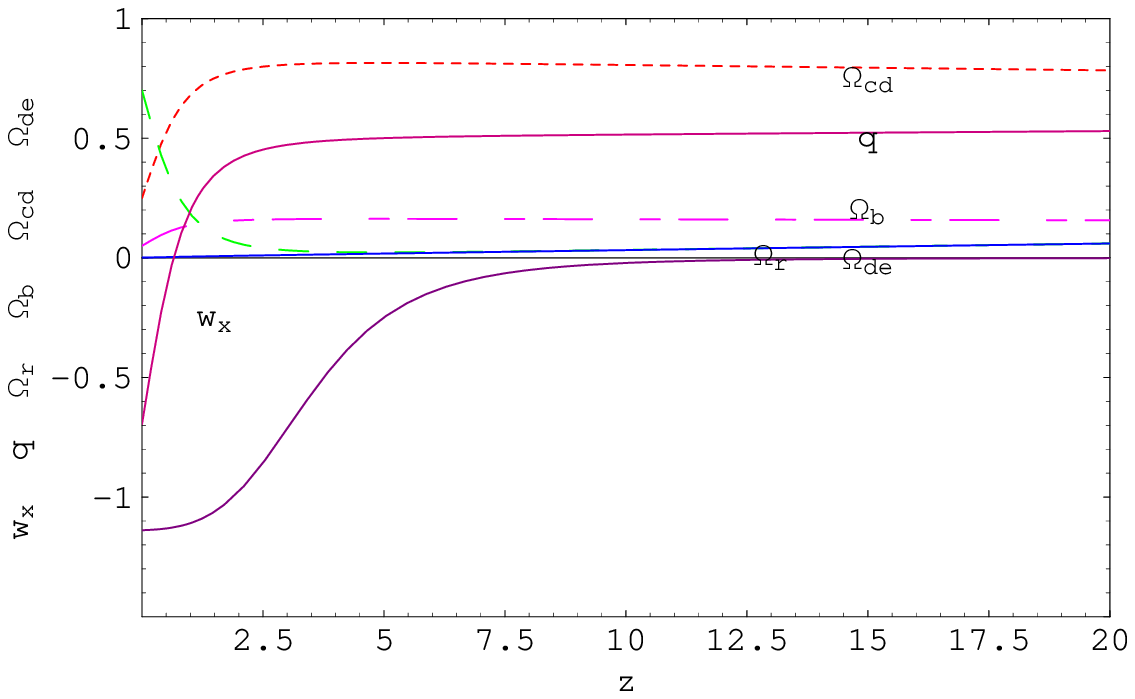}
\caption{{\bf Case I:} The evolution of the dimensionless density
parameters $\Omega_{cd}$, $\Omega_{de}$, $\Omega_{b}$,
$\Omega_{r}$ and deceleration parameter $q$, EOS of dark energy
$w_x$ versus redshift $z$, where $\Omega_{cd0}=0.25$, $\Omega_{de
0}=0.6991$, $\Omega_{b0}=0.05$, $\Omega_{b0}=0.0009$ and
$\omega_0=-1.14$.} \label{figez1}
\end{figure}

{\bf Case I}: $w_\phi=w_0$ (Ref. \cite{SHEM})
\begin{eqnarray}
\label{pot1}
\tilde{V}(z) &=& \frac{1}{2}(1-w_0)(1+z)^{3(1+w_0)},\\
E(z)&=&(1+z)^{3(1+w_0)}, \\
f(z)&=&(1+z)^2\left[\Sigma_{i}\Omega_{\gamma_{i}
0}(1+z)^{3(\gamma_{i}+1)}+\Omega_{\phi
0}(1+z)^{3(1+w_0)}\right]^{-1}.\label{fz-1}
\end{eqnarray}
The evolutions of the dimensionless energy density parameters
$\Omega_i$s, EOS of dark energy $\omega_x$ and decelerated
parameter $q$ are plotted in Fig. (\ref{figez1}) in this case.

\begin{figure}
\centering
\includegraphics[]{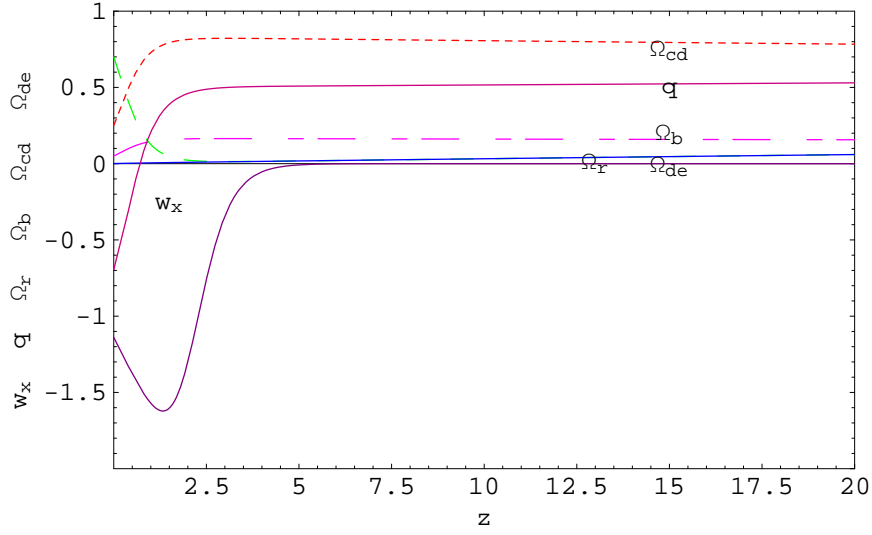}
\caption{{\bf Case II:} The evolution of the dimensionless density
parameters $\Omega_{cd}$, $\Omega_{de}$, $\Omega_{b}$,
$\Omega_{r}$ and deceleration parameter $q$, EOS of dark energy
$w_x$ versus redshift $z$, where $\Omega_{cd0}=0.25$, $\Omega_{de
0}=0.6991$, $\Omega_{b0}=0.05$, $\Omega_{b0}=0.0009$ and
$\omega_0=-1.14$ and $\omega_1=-0.5$.} \label{figez2}
\end{figure}

{\bf Case II}: $w_\phi=w_0+w_1 z$ (Ref. \cite{ARCDH})
\begin{eqnarray}
\tilde{V}(z) &=& \frac{1}{2}(1-w_0-w_1 z)
 (1+z)^{3(1+w_0-w_1)}e^{3w_1 z},\\
E(z)&=&(1+z)^{3(1+w_0-w_1)}e^{3w_1 z}, \\
f(z)&=&(1+z)^2\left[\Sigma_{i}\Omega_{\gamma_{i}
0}(1+z)^{3(\gamma_{i}+1)}+\Omega_{\phi
0}(1+z)^{3(1+w_0-w_1)}e^{3w_1 z}\right]^{-1}.\label{fz-2}
\end{eqnarray}
The evolutions of the dimensionless energy density parameters
$\Omega_i$s, EOS of dark energy $\omega_x$ and decelerated
parameter $q$ are plotted in Fig. (\ref{figez2}) for Case II.

\begin{figure}
\centering
\includegraphics[]{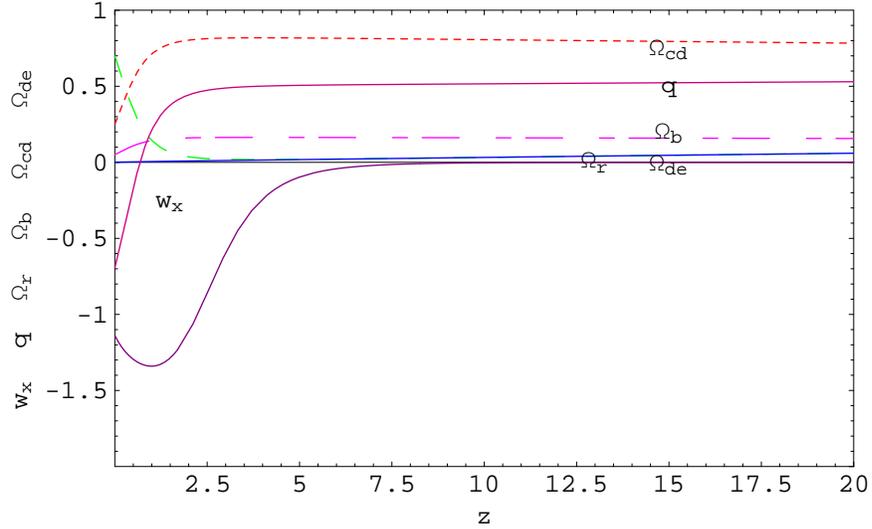}
\caption{{\bf Case III:} The evolution of the dimensionless
density parameters $\Omega_{cd}$, $\Omega_{de}$, $\Omega_{b}$,
$\Omega_{r}$ and deceleration parameter $q$, EOS of dark energy
$w_x$ versus redshift $z$, where $\Omega_{cd0}=0.25$, $\Omega_{de
0}=0.6991$, $\Omega_{b0}=0.05$, $\Omega_{b0}=0.0009$ and
$\omega_0=-1.14$ and $\omega_1=-0.5$.} \label{figez3}
\end{figure}

{\bf Case III}: $w_\phi=w_0+w_1\frac{z}{1+z}$ (Ref.
\cite{MCDP,EVL,TP})
\begin{eqnarray}
\tilde{V}(z) &=& \frac{1}{2}
 \left(1-w_0-w_1\frac{z}{1+z}\right)
 (1+z)^{3(1+w_0+w_1)}e^{-3w_1 \frac{z}{1+z}},\\
E(z)&=&(1+z)^{3(1+w_0+w_1)}e^{-3w_1 \frac{z}{1+z}},\\
f(z)&=&(1+z)^2\left[\Sigma_{i}\Omega_{\gamma_{i}
0}(1+z)^{3(\gamma_{i}+1)}+\Omega_{\phi
0}(1+z)^{3(1+w_0+w_1)}e^{-3w_1
\frac{z}{1+z}}\right]^{-1}.\label{fz-3}
\end{eqnarray}
The evolutions of the dimensionless energy density parameters
$\Omega_i$s, EOS of dark energy $\omega_x$ and decelerated
parameter $q$ are plotted in Fig. (\ref{figez3}) for Case III.

\begin{figure}
\centering
\includegraphics[]{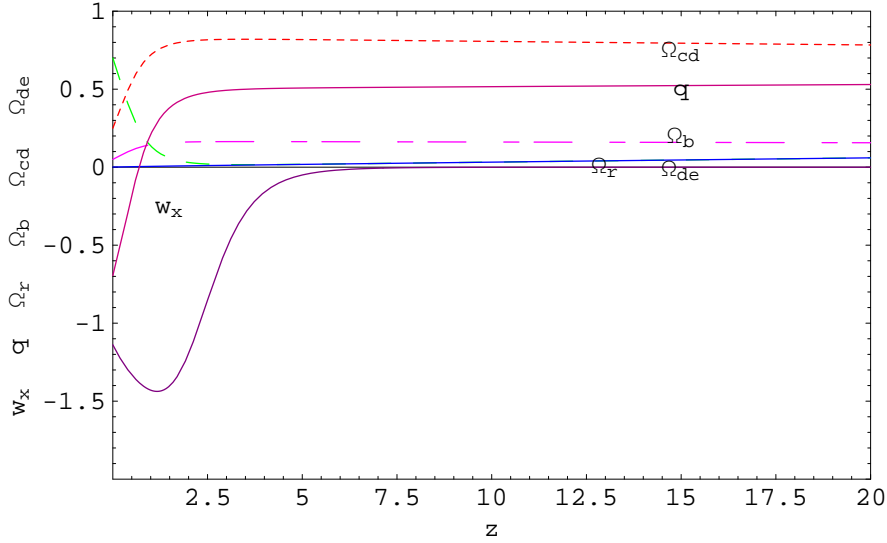}
\caption{{\bf Case IV:} The evolution of the dimensionless density
parameters $\Omega_{cd}$, $\Omega_{de}$, $\Omega_{b}$,
$\Omega_{r}$ and deceleration parameter $q$, EOS of dark energy
$w_x$ versus redshift $z$, where $\Omega_{cd0}=0.25$, $\Omega_{de
0}=0.6991$, $\Omega_{b0}=0.05$, $\Omega_{b0}=0.0009$ and
$\omega_0=-1.14$ and $\omega_1=-0.5$.} \label{figez4}
\end{figure}

{\bf Case IV}: $w_\phi=w_0+w_1\ln(1+z)$ (Ref. \cite{BFGGE})
\begin{eqnarray}
\tilde{V}(z) &=& \frac{1}{2}
 \left[1-w_0-w_1\ln(1+z)\right]
 (1+z)^{3(1+w_0)+\frac{3}{2}w_1\ln(1+z)},\\
E(z)&=&(1+z)^{3(1+w_0)+\frac{3}{2}w_1\ln(1+z)},\\
f(z)&=&(1+z)^2\left[\Sigma_{i}\Omega_{\gamma_{i}
0}(1+z)^{3(\gamma_{i}+1)}+\Omega_{\phi
0}(1+z)^{3(1+w_0)+\frac{3}{2}w_1\ln(1+z)}\right]^{-1}.\label{fz-4}
\end{eqnarray}
The evolutions of the dimensionless energy density parameters
$\Omega_i$s, EOS of dark energy $\omega_x$ and decelerated
parameter $q$ are plotted in Fig. (\ref{figez4}) for Case IV.

From the above examples, it is found that the dark energy
dominating the accelerating universe in our case can be obtained
by using the correspondence between $f(z)$ and $V(\phi)$. In the
above cases, we took the values of parameters at discretion to
describe the correspondence. The observation values of the
parameters can be determined by the data from supernovae
observations easily.

\section{Conclusions}

A general class of $5D$ cosmological models is characterized by a
big bounce as opposed to the big bang in $4D$ standard
cosmological model. This exact solution contains two arbitrary
functions $\mu(t)$ and $\nu(t)$, which are in analogy to the
different forms of the potential $V\left(\phi\right)$ in
quintessence or phantom dark energy models. Also, once the forms
of the arbitrary functions are specified, the universe evolution
will be determined. In this paper, we study these correspondence
between the arbitrary function $f(z)$ and the scalar field
potential $V(\phi)$. By these correspondences, we can define the
arbitrary form $f(z)$. And then the evolution of the universe is
determined by the correspondence. Four special cases of potentials
$V(\phi)$ are discussed. In all these cases, dark energy dominated
accelerating universe are obtained.

\section{Acknowledgments}
This work was supported by National Natural Science Foundation
(10273004) and National Basic Research Program (2003CB716300) of
P. R. China.


\section{References}

\begin{thebibliography}{99}
\bibitem{RS} A.G. Riess, et.al., {\it Observational evidence from supernovae for an
accelerating universe and a cosmological constant}, {\it Astron.
J.} {\bf 116} 1009 (1998), astro-ph/9805201; S. Perlmutter,
et.al., {\it Measurements of omega and lambda from 42
high-redshift supernovae}, {\it Astrophys. J.} {\bf 517} 565
(1999), astro-ph/9812133.

\bibitem{TKB} J.L. Tonry, et.al., {\it Cosmological Results from High-z Supernovae
}, {\it Astrophys. J.} {\bf 594} 1 (2003), astro-ph/0305008; R.A.
Knop, et.al., {\it New Constraints on $\Omega_M$,
$\Omega_\Lambda$, and w from an Independent Set of Eleven
High-Redshift Supernovae Observed with HST}, astro-ph/0309368;
B.J. Barris, et.al., {\it 23 High Redshift Supernovae from the IfA
Deep Survey: Doubling the SN Sample at z>0.7}, {\it Astrophys.J.}
{\bf 602} 571 (2004), astro-ph/0310843.

\bibitem{Riess} A.G. Riess, et.al., {\it Type Ia Supernova Discoveries
at $z>1$ From the Hubble Space Telescope: Evidence for Past
Deceleration and Constraints on Dark Energy Evolution},
astro-ph/0402512.

\bibitem{BHS} P. de Bernardis, et.al., {\it A Flat Universe from High-Resolution
Maps of the Cosmic Microwave Background Radiation}, {\it Nature}
{\bf 404} 955 (2000), astro-ph/0004404; S. Hanany, et.al., {\it
MAXIMA-1: A Measurement of the Cosmic Microwave Background
Anisotropy on angular scales of 10 arcminutes to 5 degrees}, {\it
Astrophys. J.} {\bf 545} L5 (2000), astro-ph/0005123; D.N. Spergel
et.al., {\it First Year Wilkinson Microwave Anisotropy Probe
(WMAP) Observations: Determination of Cosmological Parameters},
{\it Astrophys. J.} Supp. {\bf 148} 175 (2003), astro-ph/0302209.

\bibitem{Quintessence} I. Zlatev, L. Wang, and P.J. Steinhardt ,
\textit{Quintessence, Cosmic Coincidence, and the Cosmological
Constant}, {\it Phys. Rev. Lett.} {\bf 82} 896 (1999),
astro-ph/9807002; P.J. Steinhardt, L. Wang , I. Zlatev, {\it
Cosmological Tracking Solutions}, {\it Phys. Rev.} D {\bf 59}
123504 (1999), astro-ph/9812313; M.S. Turner , {\it Making Sense
Of The New Cosmology}, {\it Int. J. Mod. Phys.} A {\bf 17S1} 180
(2002), astro-ph/0202008; V. Sahni , {\it The Cosmological
Constant Problem and Quintessence}, {\it Class.Quant.Grav.} {\bf
19} 3435 (2002), astro-ph/0202076.

\bibitem{Phantom} R.R. Caldwell, M. Kamionkowski,
N.N. Weinberg, {\it Phantom Energy: Dark Energy with w $<-1$
Causes a Cosmic Doomsday}, {\it Phys. Rev. Lett.} {\bf 91} 071301
(2003), astro-ph/0302506; R.R. Caldwell, {\it A Phantom Menace?
Cosmological consequences of a dark energy component with
super-negative equation of state}, {\it Phys. Lett.} B {\bf 545}
23 (2002), astro-ph/9908168; P. Singh, M. Sami, N. Dadhich, {\it
Cosmological dynamics of a phantom field}, {\it Phys. Rev.} D {\bf
68} 023522 (2003), hep-th/0305110; J.G. Hao, X.Z. Li , {\it
Attractor Solution of Phantom Field}, {\it Phys.Rev.} D {\bf 67}
107303 (2003), gr-qc/0302100.

\bibitem{sahni} V. Sahni, {\it Theoretical models of dark
energy}, {\it Chaos. Soli. Frac.} {\bf 16} 527 (2003).

\bibitem{K-essence} Armend\'{a}riz-Pic\'{o}n, T. Damour, V. Mukhanov,
{\it k-Inflation}, {\it Physics Letters} B {\bf 458} 209 (1999);
M. Malquarti, E.J. Copeland , A.R. Liddle, M. Trodden, {\it A new
view of k-essence}, {\it Phys. Rev.} D {\bf 67} 123503 (2003); T.
Chiba , {\it Tracking k-essence}, {\it Phys. Rev.} D {\bf 66}
063514 (2002), astro-ph/0206298.

\bibitem{GOZ} Z.K. Guo, N. Ohtab and Y.Z. Zhang, {\it Parametrization of Quintessence
and Its Potential}, astro-ph/0505253.

\bibitem{Kaluza} T. Kaluza, {\it On The Problem Of Unity In Physics}, Sitzungsber. Preuss. Akad. Wiss.
Berlin (Math. Phys.) K1 966 (1921).

\bibitem{Klein} O. Klein, {\it Quantum Theory And Five-Dimensional Relativity}, Z. Phys. {\bf 37} 895 (1926)
[Surveys High Energ. Phys. 5 (1926) 241].

\bibitem{Compbell} J.E. Campbell, {\it A Course of Differential Geometry},
(Clarendon Oxford, 1926); S. Rippl, R. Romero, R. Tavakol, {\it
Gen. Quantum Grav.} {\bf 12} 2411 (1995); C. Romero, R. Tavako and
R. Zalaletdinov, {\it Ge. Relativ. Gravit.} {\bf 28} 365 (1996);
J. E.Lidsey, C. Romero, R. Tavakol and S. Rippl, {\it Class.
Quantum Grav.} {\bf 14} 865 (1997); S.S. Seahra, P.S. Wesson, {\it
Application of the Campbell-Magaard theorem to higher-dimensional
physics}, {\it Class. Quant. Grav.} {\bf 20} 1321 (2003),
gr-qc/0302015.

\bibitem{Wesson} P.S. Wesson, \textit{Space-Time-Matter} (Singapore: World
Scientific) 1999; J.M. Overduin and P.S. Wesson, \textit{Phys.
Rept.} {\bf 283}, 303 (1997), gr-qc/9805018.

\bibitem{FSS} V. Frolov, M. Snajdr, D. Stojkovic, {\it Interaction of a brane with a moving bulk black
hole}, {\it Phys. Rev.} D {\bf} 68, 044002 (2003).

\bibitem{LiuW} H.Y. Liu and P.S. Wesson, {\it Universe models with a variable
cosmological ``constant'' and a ``big bounce''}, {\it Astrophys.
J.} {\bf 562} 1 (2001), gr-qc/0107093.

\bibitem{STM-cosmology} T. Liko, P.S. Wesson, {\it The Big Bang as a Phase
Transition}, {\it Int. J. Mod. Phys.} A {\bf 20} 2037 (2005),
gr-qc/0310067; S.S. Seahra, P.S. Wesson, {\it Universes encircling
five-dimensional black holes}, {\it J. Math. Phys.}, {\bf 44} 5664
(2003); Ponce de Leon J, {\it Gen. Relativ. Gravit.} {\bf 20} 539
(1988); L.X. Xu , H.Y. Liu, B.L. Wang, {\it Big Bounce singularity
of a simple five-dimensional cosmological model}, {\it Chin. Phys.
Lett.} {\bf 20} 995 (2003), gr-qc/0304049; H.Y. Liu, {\it Exact
global solutions of brane universe and big bounce}, {\it Phys.
Lett.} B {\bf 560} 149 (2003), hep-th/0206198.

\bibitem{WLX} B.L. Wang, H.Y. Liu, L.X. Xu, {\it Accelerating Universe in a Big Bounce
Model}, Mod. Phys. Lett. A {\bf 19} 449 (2004), gr-qc/0304093;
L.X. Xu, H.Y. Liu, {\it Three Components Evolution in a Simple Big
Bounce Cosmological Model}, {\it Int. J. Mod. Phys.} D {\bf 14}
883 (2005), astro-ph/0412241.

\bibitem{Liu} H.Y. Liu and B. Mashhoon, {\it A machian interpretation of the
cosmological constant}, {\it Ann. Phys}. {\bf 4} 565 (1995).

\bibitem{SHEM} S. Hannestad and E. Mortsell, {\it Phys. Rev.} D {\bf 66} 063508 (2002).

\bibitem{ARCDH} A.R. Cooray and D. Huterer, {\it Astrophys. J.} {\bf 513} L95 (1999).

\bibitem{MCDP} M. Chevallier, D. Polarski, {\it Int. J. Mod. Phys.} D {\bf 10} 213
(2001); gr-qc/0009008.

\bibitem{EVL} E.V. Linder, {\it Phys. Rev. Lett.} {\bf 90} 091301 (2003).

\bibitem{TP} T. Padmanabhan and T.R. Choudhury, {\it Mon. Not. Roy. Astron. Soc.} {\bf 344} 823 (2003).

\bibitem{BFGGE} B.F. Gerke and G. Efstathiou, {\it Mon. Not. Roy. Astron. Soc.} {\bf 335} 33 (2002).

\bibitem{Ponce} J. Ponce de Leon, {\it Gen. Relativ. Gravit.} {\bf 20} 539 (1988).

\bibitem{Dahia} F. Dahia and C. Romero, {\it Dynamically generated embeddings of
spacetime}, gr-qc0503103.

\end{thebibliography}
\end{document}